\documentclass[%
 reprint,
 superscriptaddress,
 amsmath,amssymb,
 aps,
]{revtex4-2}

\usepackage{graphicx}
\usepackage{dcolumn}
\usepackage{bm}
\usepackage{makecell}
\usepackage{braket}
\usepackage{multirow, tabularx}
\newcolumntype{L}{>{\raggedright\arraybackslash}X}
\newcommand\mcl[1]{\multicolumn{2}{|l|}{#1}}

\begin{document}

\preprint{arxiv/something}
\title{Demonstration of RIP gates in a quantum processor with negligible transverse coupling}

\author{Muir Kumph}
 \email{mkumph@us.ibm.com}
\author{James Raftery}
 \email{james.raftery@ibm.com}
\author{Aaron Finck}
\author{John Blair}
\author{April Carniol}
\author{Santino Carnevale}
\author{George A Keefe}
\author{Vincent Arena}
\author{Shawn Hall}
\author{David McKay}
\author{George Stehlik}
\affiliation{IBM T.J. Watson Research Center}

\date{\today}

\begin{abstract}
Here, we report the experimental demonstration of a novel multi-mode linear bus interferometer (LBI) coupler in a six qubit superconducting quantum processor.
A key feature of this coupler is an engineered multi-path interference which eliminates transverse coupling between qubits over a wide frequency range.
This negligible static coupling is achieved without any flux bias tuning, and greatly reduces the impact of qubit frequency collisions.
We achieve good simultaneous single qubit gate operation and low $ZZ$ rates (below 600\,Hz) across the device without staggering qubit frequencies, even in cases where qubits are as close as 10\,MHz.
Multi-qubit interactions are still possible through the coupler using microwave-driven resonator induced phase gates, which we utilize to demonstrate simultaneous two qubit gates with fidelities as high as 99.4\%.



\end{abstract}

\maketitle


\section{\label{sec:intro}Introduction}

A 2D lattice of locally connected superconducting qubits has emerged as a likely candidate for building a fault tolerant quantum computer\,\cite{RevModPhys.93.025005}. In such a configuration, single qubit gates are realized by microwave driving the qubits at their resonance frequency. Two-qubit gates -- the building block for entanglement -- can be realized in several ways depending on the particular type of coupling that exists between these qubits. The simplest coupling type is a static ``always-on'' transverse coupling, which can be described by the Duffing Hamiltonian as
\begin{eqnarray}
    H_t/\hbar & = & \sum_{i} \omega^q_i a_i^\dagger a_i + \frac{\delta_i}{2} a_i^\dagger a_i (a_i^\dagger a_i -1) + \nonumber \\
    & & \sum_{ij} J_{ij}(a_i^\dagger+a_i)(a_j^\dagger+a_j) \label{eqn:duff}
\end{eqnarray}
where $\omega^q_i$ is the qubit angular frequency, $\delta_i$ is the anharmonicity of the non-linear Duffing oscillator and $J_{ij}$ is the transverse coupling strength between qubits $i$ and $j$. When $J_{ij}=0$ except for neighboring qubits, and these qubits are separated in frequency much more than their coupling $J$, this gives a dispersive interaction\,\cite{PhysRevA.79.013819}. In this configuration, two-qubit interactions can be mediated by microwave driving a control qubit at a neighbor qubit's frequency, which activates the cross-resonance interaction \cite{paraoanu:2006, rigetti:2010, chow:2011}. While this can attain errors in the low $10^{-3}$ range~\cite{PhysRevLett.127.130501}, the nature of always-on interactions means that there are constantly spectator errors when driving the gates. These errors can be particularly difficult to mitigate since they can appear non-Markovian in the qubit frame\cite{wei:2023, PhysRevA.79.013819}. Furthermore, a finite $J$ in the dispersive regime from Eqn.~\ref{eqn:duff}, leads to a finite $ZZ$ interaction term, typically written as $CZ$ rate $\xi$ between qubits $i$ and $j$ as,
\begin{eqnarray}
    H_{ij} = \xi_{ij} a_i^\dagger a_i a_j^\dagger a_j
\end{eqnarray}. 

One solution to this problem is to introduce a tunable multi-path coupler where, due to path interference, the transverse coupling between qubits is zero at a particular value of the tuning parameter\cite{PhysRevLett.127.080505, yan:2018, mundada:2019, foxen:2020},  i.e., $J=J(t)$. In this way, the coupler can be set so that there are zero spectator crosstalk errors during single qubit microwave gate operation, $J(t_0)=0$. Since $\xi \propto J^2$ this also removes any $ZZ$ interaction. Two-qubit gates can be enabled by tuning to a configuration with large interactions, $J(t_1)=J_{\textrm{int}}$. This does require an extra control knob (tunability), which may introduce new noise and instability. Furthermore, it does not solve all spectator errors as new collisions may occur as the system is tuned~\cite{PhysRevApplied.14.024042}.

In this work, we build upon the concept of interferometric cancellation and introduce a multi-mode coupler with broad-band cancellation of exchange coupling $J$, with a zero near the target qubit frequencies. Rather than depending on tunability, we dynamically disrupt the coupling interference by virtually populating one of the modes in the coupler. This action generates a phase gate between neighboring pairs~\cite{RIPpulseShapes}. During this gate the exchange coupling remains zero and there are no frequency induced qubit collisions in the device. The architecture is implemented using IBM's fixed frequency superconducting hardware with circuit QED as the foundation.  The device is described below in section\,\ref{subsec:layout} including a derivation of the multi-mode Hamiltonian and resonator phase interaction.  Next, the testing and performance of the device are shown in section\,\ref{sec:testing}.  And finally a discussion and outlook are given in section\,\ref{sec:discussion}.

\section{\label{subsec:layout}Experimental Device}

\begin{figure}
    \centering
    \includegraphics[width=0.9\columnwidth]{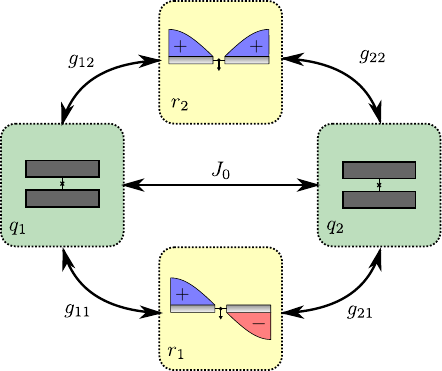}
    \caption{Each neighboring qubit $i$ is coupled to resonator mode $l$ with an exchange coupling $g_{il}$.  In addition there is a direct exchange coupling $J_0$ between the pair of qubits.  By selecting the appropriate couplings $g_{il}$ and $J_0$ a broadband cancellation with a zero in coupling can be made near the typical qubit frequencies.}
    \label{fig:coupling}
\end{figure}

The test device consists of 6 qubits arranged in a square lattice. Between each pair of qubits there is a multi-mode coupler as shown in figure\,\ref{fig:coupling}, which allows for phase gates to be driven between each pair of qubits while keeping the exchange coupling $J$ between pairs to a negligible value.  When the multi-mode coupler is in the ground state, the negligible value of $J$ means that the static ZZ interaction between pairs is negligible as well.  This allows for a lattice of qubits where all interactions are off except when the multi-mode coupler is driven.

The device described above has six qubits connected on a square lattice.  In this section, just a two qubit and single coupler subsystem is considered.  The coupler has two modes with a broadband cancellation and design-adjustable zero in coupling $J$.  Figure\,\ref{fig:coupling} shows the overall coupling and the synthesis of this coupling of the multi-mode coupler.  Each neighboring qubit $i$ is coupled to resonator mode $l$ with a signed exchange coupling $g_{il}$.  In addition there is a direct exchange coupling $J_0$ between the pair of qubits.  By selecting the appropriate values of couplings $g_{il}$ and $J_0$, a broadband cancellation with a zero in coupling can be made near the typical qubit frequencies.  Next the Hamiltonian of this subsystem and effective coupling $J$ is derived in subsection\,\ref{sec:couplerHam} and then the superconducting circuit is described in subsection\,\ref{subsec:sclayout}.

\subsection{\label{sec:couplerHam}Hamiltonian of the LBI coupler}

The key features of the LBI coupler can be explored using a simplified model of two transmon qubits \cite{transmon} coupled to two linear oscillator modes.  
Here we extend the block diagonal analysis performed in \cite{RIPpulseShapes}, to one of 2 qubits, with direct coupling to each other and coupling to 2 resonator modes. While this analysis doesn't include higher order effects\,\cite{PhysRevA.109.012601}, this treatment does inform the basic principles of the device operation. The Hamiltonian for the LBI coupler can be modeled as follows:
\begin{equation}
\begin{split}
    H/\hbar=\sum_i H_t(i) + \sum_l H_c(l) + \sum_{il} H_{int}(i,l) 
\end{split}
\end{equation}
where $i,l \in \{1,2\}$ correspond to the qubit number and bus resonator mode.
As stated in the introduction, the two charge qubits in the transmon regime can be modeled using a qudit Duffing Hamiltonian as
\begin{equation}
    H_t(i)/\hbar = \omega^q_i a_i^\dagger a_i + \frac{\delta_i}{2} a_i^\dagger a_i (a_i^\dagger a_i -1)
\end{equation}
where $\omega^q_i$ is the qubit angular frequency and $\delta_i$ is the anharmonicity of the non-linear Duffing oscillator.
The cavity Hamiltonian is
\begin{equation}
    H_c(l)/\hbar = \omega^r_l c_l^\dagger c_l.
\end{equation}
where $\omega^q_l$ is the cavity mode angular frequency. 
Lastly, the interaction Hamiltonian is
\begin{equation}
    H_{int}(i,l)/\hbar = \sum_{il} g_{il} (a_i c_l^\dagger + a_i^\dagger c_l) + J_0(a_1 a_2^\dagger + a_1^\dagger a_2)
\end{equation}
where the exchange coupling between qubit $i$ and resonator mode $l$ is $g_{il}$ and the direct exchange coupling between qubits 1 and 2 is given by $J_0$.

In the dispersive regime, where the detuning $\Delta_{il}$ between qubit $i$ and resonator mode $l$ is much greater than the coupling $g_{il}$, the dispersive Hamiltonian is approximately
\begin{equation}
\begin{split}
    H_d/\hbar =  \sum_l H_1^{\prime}(\ket{m}, c_l) + \sum_l H_2^{\prime}(\ket{n}, c_l) +\\
    \sum_l H_c(l) + H^{\prime}_{int}
\end{split}
\end{equation}
where the resonator frequency shift due to the interaction has been neglected.  The qudit Hamiltonian in the dispersive limit 
\begin{equation}
    H^\prime_i(\ket{k}, c_l)/\hbar = \tilde{\omega}^q_i \ket{k} \bra{k} + \sum_{kl} \chi_{i,k,l} c_l^\dagger c_l \ket{k} \bra{k} 
\end{equation}
with 
\begin{equation}
\begin{split}
\tilde{\omega}_i=k\omega^q_i + \frac{\delta_i}{2} k (k -1) + \sum_l \frac{k g_{il}^2}{\omega^q_i + (k-1)\delta_i - \omega^r_l}\\
\chi_{i,k,l} = \frac{g^2_{il}(\delta_i - \omega^q_i + \omega^r_l)}{(\omega^q_i + k\delta_i - \omega^r_l)(\omega^q_i + (k-1)\delta_i - \omega^r_l)}.
\end{split}
\label{eqn:dispShift}
\end{equation}
For qubit $i$ in eigenstate $\ket{k}$ interacting with bus mode $l$, the effective interaction Hamiltonian is then
\begin{equation}
\begin{split}
    H^{\prime}_{int}/\hbar=\sum_{j,k,l}\sqrt{(j+1)(k+1)}J_{j,k,l}(\ket{j,k+1}\bra{j+1,k} \\ + \ket{j+1,k}\bra{j,k+1})\\
    +\sum_{j,k}\sqrt{(j+1)(k+1)}J_0(\ket{j,k+1}\bra{j+1,k} \\ + \ket{j+1,k}\bra{j,k+1})
\end{split}
\end{equation}
where the effective coupling for bus mode $l$
\begin{equation}
    J_{j,k,l}=\frac{g_{1l} g_{2l} (\omega^q_1 + \omega^q_2 + j\delta_1 + k\delta_2 - 2\omega^r_l)}{2(\omega^q_1+j\delta_1-\omega^r_l)(\omega^q_2+k\delta_2-\omega^r_l)}.
\end{equation}
If effective coupling is only considered in the qubit subspace, then the effective coupling is
\begin{equation}
    J_l=\frac{g_{1l} g_{2l}(\omega^q_1+\omega^q_2 - 2\omega^r_l)}{2(\omega^q_1 - \omega^r_l)(\omega^q_2 - \omega^r_l)}.
\end{equation}
The total effective coupling $J$ can then be approximated as a sum of the coupling from bus modes 1,2 and the direct coupling $J_0$
\begin{equation}
    J=J_0 + \sum_{l} J_l.
\label{eqn:Jtotal}
\end{equation}
The total coupling can be engineered to be zero because the sign of $J_l$ depends on the parity of the qubit-bus mode couplings $g_{il}$.  Furthermore, the individual couplings $J_l$ and total coupling $J$ are independent of the bus mode excitation.  This suggests that errors and/or collisions due to exchange coupling can be suppressed even while energizing the bus.

In contrast, the ZZ rate between qubits is dependent on the bus mode excitation. If, however, the buses are in their ground state, the ZZ coupling rate $\zeta_0$ between qubits is related to the total coupling $J$ and can be approxmated as
\begin{equation}
    \zeta_0 = - \frac{2J^2(\delta_1+\delta_2)}{(\delta_1 + \Delta_{12})(\delta_2 - \Delta_{12})}
\end{equation}
where $\Delta_{12}=\omega^q_1 - \omega^q_2$ is the qubit detuning.
This allows for both the total coupling $J$ and the static qubit-qubit ZZ coupling to be made zero by engineering the appropriate qubit-bus mode couplings $g_{il}$ and a direct coupling $J_0$.

\begin{figure}[hbt!]
    \centering
    \includegraphics[width=\columnwidth]{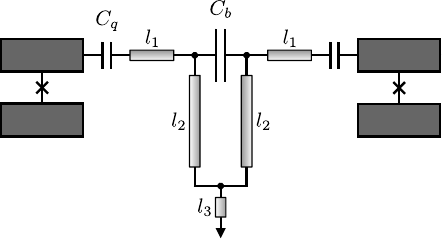}
    \caption{Circuit schematic of the LBI coupler connecting two transmon qubits. Each qubit is strongly coupled to the bus through the $C_q$ capacitors, and to each other through bypass capacitor $C_b$. Two degenerate $\lambda/4$ resonators are formed via CPW sections $l_1$, $l_2$, and shared section $l_3$. Coupling between the two resonators, determined by $C_b$ and $l_3$, gives rise to symmetric and anti-symmetric $\lambda/2$ modes which couple to both qubits but cancel one another due to their opposite polarities.}
    \label{fig:couplercircuit}
\end{figure}

Critically, the dispersive shift of the qubit due to the bus modes is independent of the sign of the qubit-bus mode couplings $g_{il}$. This provides an avenue for functional two-qubit resonator induced phase (RIP) gates\,\cite{RIPgates} even if the total effective coupling $J$ is negligible. In this paper we explore a parameter regime where we have a strong qubit-bus dispersive shift of $2\chi=\mathcal{O}(1)$\,MHz while still maintaining a small qubit-qubit coupling of $J\simeq0.1$\,MHz and $\zeta_0 < 1$\,kHz. 


\subsection{\label{subsec:sclayout}Superconducting Circuit Implementation}
In order to realize the above coupling scheme, an LBI bus was designed with two degenerate modes coupled together\,\cite{LBIPatent}.  The two modes then form an odd and even mode which allow for substantial cancellation over a broad frequency range.  Introducing a small bypass capacitor to cancel the residual coupling allows for a true zero in coupling to be placed near the qubit frequencies. Figure\,\ref{fig:couplercircuit} shows the schematic of such a circuit.  Transmon style qubits on the left and right are connected with a co-planar waveguide (CPW) to a bypass capacitor in the center.  This bypass capacitor provides direct coupling $J_0$.  T-junctions provide for the CPW to be continued down so that the circuit forms two $\lambda/4$ resonators with the same frequency.  A shared stub at the bottom of the circuit provides coupling between the modes of the two buses.  The stub and bypass capacitor allow for tuning the frequencies of both the symmetric $\lambda/4$-like mode and the anti-symmetric $\lambda/2$-like mode formed by the hybridization of the two $\lambda/4$ resonators.  
The bus modes have opposite electrodynamic parity so that the effective qubit-qubit coupling $J_{l}$ is opposite for each mode.  
Microwave simulations\,\cite{firat2019} were used to tune the bus stub length and bypass capacitance so that the effective total coupling $J$ is zero, while maintaining an appreciable qubit-bus dispersive shift $2\chi$. 

\begin{figure}[hbt!]
    \centering
    \includegraphics[width=\columnwidth]{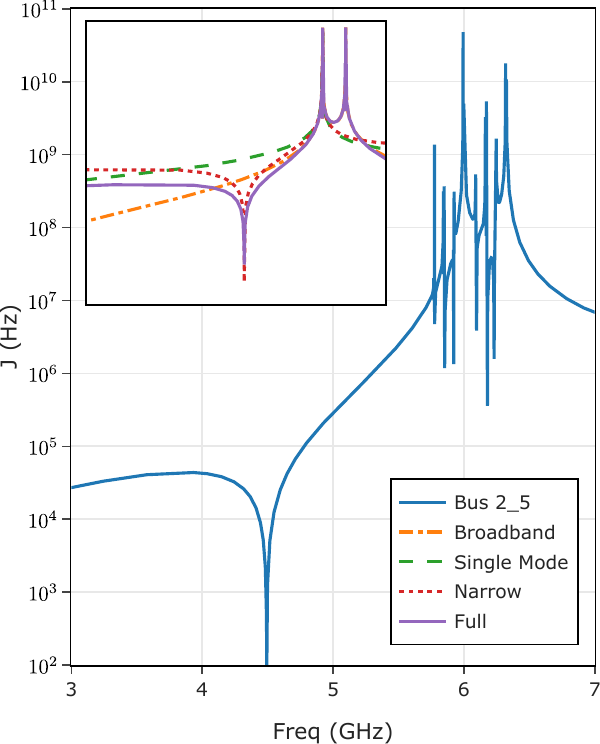}
    \caption{Calculated exchange coupling $J$ as a function of qubit frequency (degenerate qubits) for bus 2\_5 using impedance response formulas\cite{firat2019} and microwave simulations.  The device was designed with a zero in coupling at roughly 4.5\,GHz.  The full device sims capture the resonances of the other bus modes, however the dynamics of the modes for bus 2\_5 are dominant in the range of qubit frequencies 4-5.5\,GHz.  The inset compares this multi-mode bus response compared to other common schemes (see main text).}
    \label{fig:coupling_sim}
\end{figure}


The inset of figure\,\ref{fig:coupling_sim} illustrates the qualitative features of the LBI coupler by comparing the total coupling strength versus frequency for similar coupling schemes as seen in the literature.
The simplest is the \emph{Single Mode} bus coupler, characterized by a single resonance peak and a broad region of large coupling below resonance.
As discussed above, the large static coupling leads to unwanted errors and spectator collisions and idle ZZ errors.
An alternative \emph{Narrow} cancellation scheme \cite{PhysRevLett.127.130501} makes use of a bypass capacitor in parallel with a single mode resonator.  
With this design a zero in both J and static ZZ can be placed at the nominal qubit frequency, however the narrowband nature of the cancellation effect makes it challenging to leverage.
Using two resonator modes one can achieve a Broadband cancellation where the positive and negative couplings of the two modes compete, leaving only a small residual coupling over a large frequency range.
The LBI bus combines the frequency dependence of the two mode bus with the tunable zero provided by the bypass capacitor to create a broadband zero in both transverse coupling and static ZZ.
A microwave simulation of the full six qubit processor (main panel) shows good agreement with this intuition.
Here the total exchange coupling between qubits 2 and 5 is shown as a function of frequency.
Parameters were chosen such that the zero in coupling occurs at 4.5\,GHz.

\subsection{\label{subsec:ripGates}RIP Gates}
The device as described provides some broadband cancellation of coupling $J$ between qubits.  This can avoid unwanted couplings that can give rise to spectator errors and collisions.  The significant qubit-bus dispersive shift of the circuit allows for a controlled phase gate in the form of a resonator induced phase (RIP) gate\,\cite{RIPgates}.  By energizing mode(s) of the coupler, a ZZ interaction is turned on between qubits connected to the coupler.  The RIP gate drive is applied to the linear resonator between the 2 qubits. In order to excite the resonator without causing measurement induced dephasing, the gate drive is applied at a frequency $\omega_d$ with a detuning $\Delta=\omega_d-\omega^r_l$.  The bandwidth is inversely proportional to the gate time $t_g$.  Generally, the detuning is chosen to be larger than the bandwidth of the pulse, which allows for an adiabatic turn-on and turn-off of the ZZ-coupling to avoid leaving energy in the resonator\,\cite{RIPpulseShapes}. For degenerate qubits, the RIP gate speed is given by\,\cite{RIPgates}
\begin{equation}
    \dot{\theta}_{ZZ}=-\frac{|\epsilon_0|^2\chi^2}{8\Delta(\Delta+2\chi)(\Delta+4\chi)}
\end{equation}
where $\epsilon_0$ is the strength of the drive applied to the resonator and $2\chi$ is the dispersive shift between the qubits and the resonator mode being used for the gate. 

One feature of the RIP gate is that its power scales strongly with reduced gate speed.  The detuning $\Delta$ must increase with the bandwidth of the drive pulse (inverse of the gate time) to avoid populating the coupling resonator.  And the gate speed also needs to increase linearly as the gate time is reduced. This means that the power of the RIP drive ($P\propto|\epsilon|^2_0$) scales with the 4th power of the inverse of the gate time.  This steep increase in drive power is the cause of the demise of the RIP gate in most modern superconducting quantum experiments as it strictly limits how short one can make the gate before multi-photon non-linear processes dominate\,\cite{PhysRevA.105.022607}.  

\section{\label{sec:testing}Testing and Results}

\begin{figure}
    \centering
    \includegraphics[width=\columnwidth]{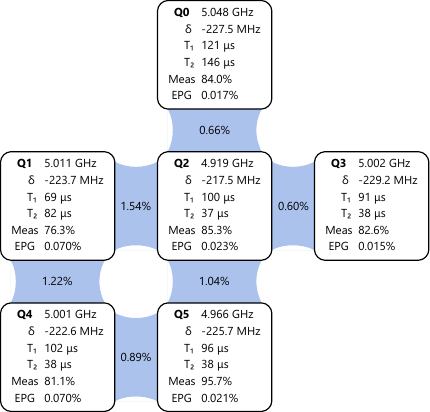}
    \caption{
        A schematic map of the measured device consisting of 6 qubits connected by LBI couplers.
        Individual qubit details (boxes) include the transition frequency ${\omega}/{2\pi}$ , anharmonicity $\delta$, energy relaxation time $T_1$, dephasing time $T_2$, measurement assignment fidelity, and simultaneous single qubit error per gate EPG.
        Blue connections between qubits show two-qubit randomized benchmarking error per gate.
    }
    \label{fig:device}
\end{figure}

\begin{figure}
    \centering
    \includegraphics[width=\columnwidth]{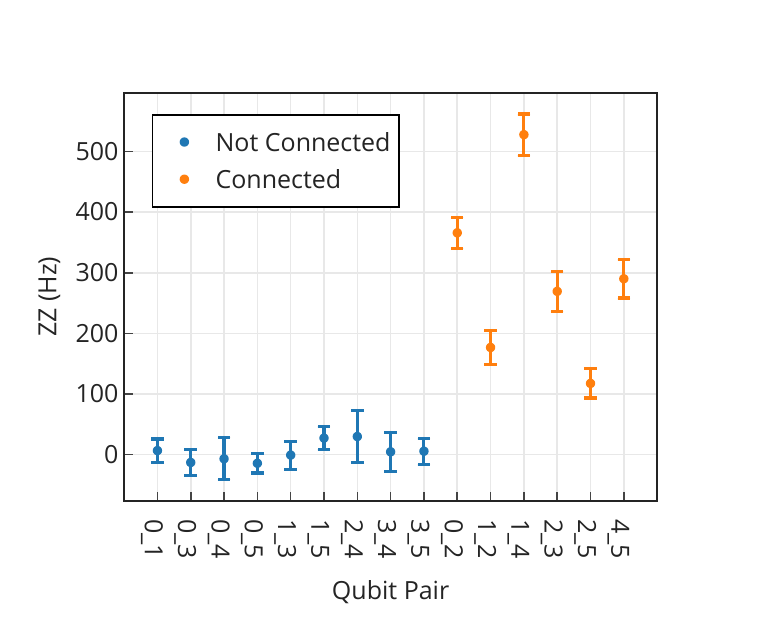}
    \caption{Measured static ZZ between all pairs of qubits.}
    \label{fig:zz}
\end{figure}

The device parameters are shown in figure\,\ref{fig:device}. It shows a schematic map of the 6 qubit LBI bus device.  Each qubit is shown with its transition frequency, anharmonicity, coherence properties, measurement assignment fidelity and single-qubit-gate error rate from simultaneous randomized benchmarking~\cite{gambetta:2012}.  The connections between pairs show the error rate of two-qubit ZZ gates.   Gate fidelity was optimized by varying the detuning and gate length for each qubit pair. In figure\,\ref{fig:zz} the measurements of static ZZ interactions between the qubits are shown. 

\begin{table}[hbt!]
    \renewcommand\tabularxcolumn[1]{m{#1}}
    \sffamily
    \renewcommand\arraystretch{1.2}
    \centering
    
\begin{tabularx}{0.8\columnwidth}{|*{7}{l|}l|}
  \cline{1-8}
\mcl{Bus ($i\_j$)} & 0\_2 & 1\_2 & 1\_4 & 2\_3 & 2\_5 & 4\_5 \\
  \cline{1-8}
\multirow{2}{8mm}{Mode [GHz]}
    & 1 & 5.795 & 5.942 & 5.794 & 5.868 & 6.017 & 5.870 \\
  \cline{2-8}
    & 2 & 6.087 & 6.233 & 6.071 & 6.163 & 6.310 & 6.159 \\     
  \cline{1-8}
\end{tabularx}
    \caption{
        Basic properties of the bus resonators.
        The first row indicates the bus between each pair of connected qubits (e.g. bus $i\_j$ connects qubits $i$ and $j$).  
        The second row indicates the frequency of each mode $l$ for that bus.
        }
    \label{tab:resprops}
\end{table}
Table~\ref{tab:resprops} lists basic resonator properties of each bus.  The device was designed with 4 pairs of bus frequencies so that bus 1\_4/4\_5  has the same nominal frequencies as 0\_2/2\_3 (accordingly).  

Table~\ref{tab:res12_25} lists more detailed resonator properties including the coupling of the qubits to these resonators.  The dispersive interactions between qubits and resonators were measured with number splitting spectroscopy.  Because of the nearby resonances, the dispersive shift as measured by number splitting spectroscopy of the degenerate buses (1\_4/0\_2 ) and (2\_3/4\_5) was hard to interpret.  And our simplistic treatment of a 2 qubit - one bus model isn't valid.  For bus 1\_2 and 2\_5 properties are given, where the coupling of each resonator mode to each qubit is calculated from measured properties using equation\,\ref{eqn:dispShift}.

\begin{table}[b]
 \renewcommand\tabularxcolumn[1]{m{#1}}
    \sffamily
    \renewcommand\arraystretch{1.2}
    \centering
\begin{tabularx}{1\columnwidth}{|c|c|c|c|c|r|r|r|}
\cline{1-8}
\thead[cc]{Bus \\ ($i\_j$)}&
\thead[cc]{$J_0$ \\ \textrm{[MHz]}}&
\thead[cc]{Mode \\ ($l$)}&
\thead[cc]{\textrm{Freq} \\ \textrm{[GHz]}}&
\thead[cc]{$2\chi_{il}$ \\ \textrm{[MHz]}}&
\thead[cc]{$2\chi_{jl}$ \\ \textrm{[MHz]}}&
\thead[cc]{$g_{il}$ \\ \textrm{[MHz]}}&
\thead[cc]{$g_{jl}$ \\ \textrm{[MHz]}}\\
\cline{1-8}
$1\_2$ & 3.0 & 1 & 5.942 & 2.5 & 2.1 & $77\,\pm\,4$ & $-78\,\pm\,5$\\
\cline{3-8}
		& & 2 & 6.233 & 2.7 & 2.7 & $103\,\pm\,5$ & $112\,\pm\,5$\\
\cline{1-8}
$2\_5$ & 2.8 & 1 & 6.017 & 2.0 & 2.0 & $81\,\pm\,5$ & $-77\,\pm\,5$\\
\cline{3-8}
		& & 2 & 6.310 & 2.3 & 2.5 & $109\,\pm\,6$ & $108\,\pm\,5$\\
\cline{1-8}
\end{tabularx}
\caption{\label{tab:res12_25}
Selected bus properties of the measured device.
The first column indicates the bus between each pair of connected qubits (e.g. bus $i\_j$ connects qubits $i$ and $j$).  
The second column indicates the inferred direct exchange coupling term $J_0$.
The rows then split out showing the properties of each mode ($l$).
The frequency of the mode is shown in the 4th column.
The dispersive shifts ($2\chi$) between the mode and qubits $i$ and $j$ measured using number splitting spectroscopy are given in column 5 and 6 respectively.
Column 7 and 8 give estimates of the coupling strength between the mode and qubits $i$ and $j$ based on the measured dispersive shifts.
}
\end{table}

The ZZ interactions were used to calculate a net coupling $J$ and that could then be used to back-out the static coupling $J_0$ in table\,\ref{tab:res12_25} using equation\,\ref{eqn:Jtotal}.

\begin{figure}
    \centering
    \includegraphics[width=\columnwidth]{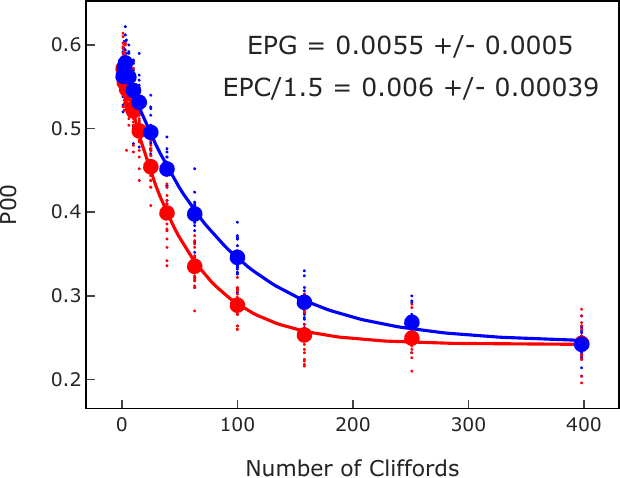}
    \caption{Two qubit interleaved randomized benchmarking results for qubit pair $2\_3$. The gate length is 228 ns with an estimated coherence limited gate fidelity of 99.4\%.  The probability of the qubits returning to the $\ket{00}$ state is shown as a function of Clifford sequence length. Fitting both decay curves we measure an EPG of 0.0055 from interleaved and see good agreement with the EPG inferred from the non-interleaved reference data (blue) as described in the main text.}
    \label{fig:gateRB}
\end{figure}

A typically good 2 qubit benchmarking result for qubit pair 2\_3 is shown in figure\,\ref{fig:gateRB}.  The gate length is 228 ns with an estimated coherence limited gate fidelity of 99.4\%.  The probability of the qubits returning to the $\ket{00}$ state is shown as a function of Clifford sequence length. Fitting the decay curves of interleaved randomized benchmarking (red) and reference randomized benchmarking (blue) we infer an EPG of 0.0055~\cite{magesan:2012}. We can upper bound the gate error from the reference curve only as $EPG \approx EPC/1.5$ since there are 1.5 two-qubit $CZ$ gates per Clifford. We see good agreement with the $EPG$ measured in both ways.

The lack of exchange coupling $J$ on this device is demonstrated by the qubit pair $1\_4$, with just 10\,MHz of detuning.  With a tunable coupling $J$ scheme, the two qubits could hybridize as their coupling was turned on, making this pair difficult to control.   For a fixed always on coupling $J\simeq 1$\,MHz this would cause a collision impacting even the performance of single qubit gates.  Their ZZ coupling rate of 500 Hz implies a residual coupling $J\simeq100$\,kHz that is small enough to avoid a collision even for detunings as low as 1\,MHz.

Typically, the RIP drive detuning was about 30\,MHz.  In order to obtain these gate fidelities, the RIP drive signals were filtered with a combination of a 100 MHz tunable band pass YIG filter (Microlambda MLFP-1742PA) in series with a 70 MHz band reject YIG filter (Micro Lambda MLRF-0408PA).  The band pass filter allowed for mixer products and the local oscillator signals to be removed from the signal generation.  And the band-reject filter lowered the noise floor at the bus frequency to avoid measurement induced dephasing of the connected qubits.

\section{\label{sec:discussion}Discussion and Outlook}

This device is a demonstration where both exchange interactions $J$ and dispersive shifts $\zeta_0$ between the qubits have been largely eliminated.  This shows one possible approach to avoiding frequency crowding on highly connected devices in quantum computing.  Furthermore, the RF driven gate here, was able to perform across a wide range of qubit-qubit detunings.  And despite the frequency crowding of the bus coupler modes, the two-qubit gates did not suffer appreciably.  A more detailed analysis of spectator errors may be warranted.  A less complex issue would be how to improve the gate speed to make the error rates feasible for use with error correcting codes.

While the RIP gate demonstrated on this device was limited in its speed, the basic physics might also be applied using a a non-linear coupling element so that the off-resonant and high-power requirement of the RIP gate is removed.  Further work using non-linear LBI type couplers may allows for faster and higher fidelity gates.

\begin{acknowledgments}
We wish to acknowledge the support of the entire IBM Quantum team for their help in building testing and developing this device.  We would like to thank Hanhee Paik, Moein Malekakhlagh and Andrew Cross especially for discussions about RIP gates.
\end{acknowledgments}

%

\end{document}